\documentclass[journal]{IEEEtran}

\usepackage[pdftex]{graphicx}
\graphicspath{{../pdf/}{../jpeg/}}
\usepackage[cmex10]{amsmath}
\usepackage{amsfonts}
\usepackage{amsthm}
\usepackage[english]{babel}
\usepackage{graphicx}
\usepackage{cite}
\usepackage{array} 
\usepackage{tabularx}
\usepackage{subcaption}
\newtheorem{theorem}{Theorem}
\usepackage{amssymb}

\begin{document}
	\title{Resource Allocation of Dual-Hop VLC/RF Systems with Light Energy Harvesting}

	\author{Shayan Zargari,~Mehrdad Kolivand,~S.Alireza Nezamalhosseini,~Bahman Abolhassani,~Lawrence R. Chen,~\textit{Senior Member,~IEEE,} and Mohammad Hossein Kahaei
		\thanks{Shayan Zargari, Mehrdad Kolivand, S.Alireza Nezamalhosseini, Bahman Abolhassani, and Mohammad Hossein Kahaei are with the School of Electrical Engineering, Iran University of Science and Technology, Tehran, Iran. L. R. Chen is with the Department of Electrical and Computer Engineering,~McGill University,~Montreal, QC H3A0G4, Canada,~e-mails: s\_zargari@elec.iust.ac.ir, mehrdad\_kolivand@elec.iust.ac.ir, nezam@iust.ac.ir, abolhassani@iust.ac.ir, lawrence.chen@mcgill.ca, and kahaei@iust.ac.ir.}}

	\maketitle
	
	\begin{abstract}
		In this paper,~we study the time allocation optimization problem to maximize the sum throughput in a dual-hop heterogeneous visible light communication (VLC)/radio frequency (RF) communication system.~Two scenarios are investigated in this paper.~For the first scenario,~we consider an optical wireless powered communication network (WPCN) in which all users harvest energy from the received lightwave over downlink (DL),~and then they use the harvested energy to transmit information signals in the uplink (UL) channels based on the time division multiple access (TDMA) scheme.~The optimal time allocation in the UL is obtained to maximize the sum throughput of all users.~For the second scenario,~the time-switching simultaneous lightwave information and power transfer (TS-SLIPT) based on the dual-hop VLC/RF is assumed that the LED transmits information and power simultaneously in the first-hop DL (i.e., VLC link).~The harvested energy at the relay is used to transmit information signals over the UL in the second-hop (i.e., RF link).~We propose a multi-objective optimization problem (MOOP) to study the trade-off between UL and DL sum-rate maximization.~The non-convex MOOP framework is then transformed into an equivalent form,~which yields a set of Pareto optimal resource allocation policies.~We also illustrate the effectiveness of the proposed approaches through numerical results.
	\end{abstract}
	

	\section{Introduction}
	Recently,~visible light communication (VLC) has attracted much attention for indoor wireless data transmission as it is capable of providing high data rates with high security and low-cost \cite{Komine}.~Besides,~as compared to the traditional radio frequency (RF) wireless communication,~VLC has many advantages such as wide unregulated bandwidth,~no RF radiation,~and no negative effects on human health \cite{Arnon}.~In particular,~intensity modulation with direct detection (IM/DD) is exploited in the VLC systems for low-cost implementation.~In this regard,~a LED is used as an optical source to convert the electrical signals into the optical signals,~and photodiodes (PDs) are employed to convert the received optical power into the electrical current at the receiver \cite{Jovicic}.~Despite the aforementioned advantages of VLC technology,~the coverage area of the VLC-based networks is small since the originated light from a LED source is confined to a limited area \cite{Elgala}.~To address this,~a dual-hop hybrid VLC/RF system has been proposed to extend the communication coverage \cite{Haas}.~In this approach,~a second-hop RF link is utilized to extend the coverage of a first-hop VLC link by introducing a relay between these two hops \cite{Rakia2}.

	However,~several applications of optical wireless networks such as the Internet of Things (IoT),~wireless sensor networks (WSNs),~and wireless personal networks (WPNs) are constrained by the finite battery capacity of the involved devices.~Energy harvesting (EH) can be regarded as an intriguing technique to substantially prolong the lifespan of batteries \cite{Khalili}.~Accordingly,~each device can harvest energy from the received signals or ambient sources (e.g., sunlight,~wind, thermal gradients) to exploit it for either signal processing purposes or communication.~More specifically,~wireless power transmission (WPT) from the light source can be obtained by the solar panel and PDs.~In \cite{Wang},~for the first time,~a solar panel was proposed as a PD to harvest energy from the DC component of the modulated light.~In particular,~it was revealed that the employment of solar panel could meet the demands of communication and EH simultaneously.~Especially,~Two schemes can be considered in EH systems combined with power transmission in optical wireless networks,~namely,~wireless powered communication networks (WPCN) \cite{Fakidis} and simultaneous lightwave information and power transfer (SLIPT) \cite{Wang1}.

	The principle of the WPCN is that receivers are first powered by the received signals in the downlink (DL) and then transmitters adopted that amount of the EH to perform wireless information transmission (WIT) in the uplink (UL) \cite{WPCN1}.~From this point of view,~in \cite{Rakia1} and \cite{Rakia2},~a dual-hop hybrid VLC/RF system was studied with the presence of a relay to extend the coverage range.~Especially,~the relay nodes harvest energy from the direct current (DC) component of the received optical signal in the first-hop DL (VLC link) and then,~they apply that for transmission data in the second-hop over the RF channels.~In particular,~the data rate maximization problem and packet failure probability was analyzed in \cite{Rakia1} and \cite{Rakia2},~respectively.

	In the SLIPT system that was first introduced in \cite{Diamantoulakis}, receivers divide the DL received signals into two streams with different levels of powers, each of which capable of information decoding (ID) and EH. Two fundamental receiver architectures have been proposed for the SLIPT systems, i.e., power splitting (PS) and time-switching (TS) \cite{Diamantoulakis}.~For the PS-based receivers,~an optimized PS ratio is used to distinguish between ID and EH,~while for the TS-based ones a given time interval is divided into two sections for ID and EH \cite{Pan}.~The deployment of a solar panel receiver with the capability of simultaneous EH and detecting signals was proposed in \cite{Min}.~The authors in \cite{Sandalidis} studied the PS-based SLIPT VLC system,~where they only investigated the hardware design and did not take any optimization of the PS ratio into account.~The DL sum-rate maximization problem in a SLIPT-based VLC system was studied in \cite{Huang}.~The authors in \cite{Ma} considered the implementation of a solar panel and a PD as the energy collector and the information decoder,~respectively.~Specifically,~the minimization of total transmit power and the maximization of minimum rate was studied.

	In this paper,~we consider two distinct scenarios in an optical WPT system.~In the first scenario,~we study the optical WPCN in which each user harvest energy from the received lightwave continuously in all time slots over the DL and then use that for transmitting information signals in the UL such that provide sufficient energy for transmission over the RF links based on the time division multiple access (TDMA) scheme.~In the second scenario,~unlike the first scenario,~the SLIPT-based dual-hop VLC/RF is assumed,~which LED transmits information and power simultaneously in the form of lightwave over the DL using the TS-based SLIPT.~In the DL,~a portion of time is dedicated to each user in order to decode information,~while the left one is utilized for EH.~More specifically,~DC and AC components are exploited for EH and ID at each receiver,~respectively,~by adopting an inductor and a capacitor \cite{Wang1}.~In each receiver,~the harvested energy is then applied to convey data in the UL over the RF links.~
	
	The contributions of this paper can be summarized as follows:
	
	• {For the first scenario,~we formulated the UL sum-rate maximization problem in a WPT-based VLC system by optimizing UL time allocations.~Although the optimization problem is non-convex,~it can be converted into an equivalent convex one which yields an optimal solution.}
	
	• For the second scenario,~we introduced a MOOP framework,~which aims at maximizing the UL and DL sum-rates simultaneously by jointly optimizing the transmission power of each user as well as UL and DL time allocations.~The MOOP can provide an appealing trade-off between the EH and ID in the SLIPT-assisted VLC system,~where the UL and DL sum-rates will be affected accordingly.
	
	• Finally,~simulation results are revealed to compare and validate WPT and SLIPT performances in the dual-hop VLC/RF system.

	\begin{figure}
		\centering
		\includegraphics[width=0.42\textwidth]{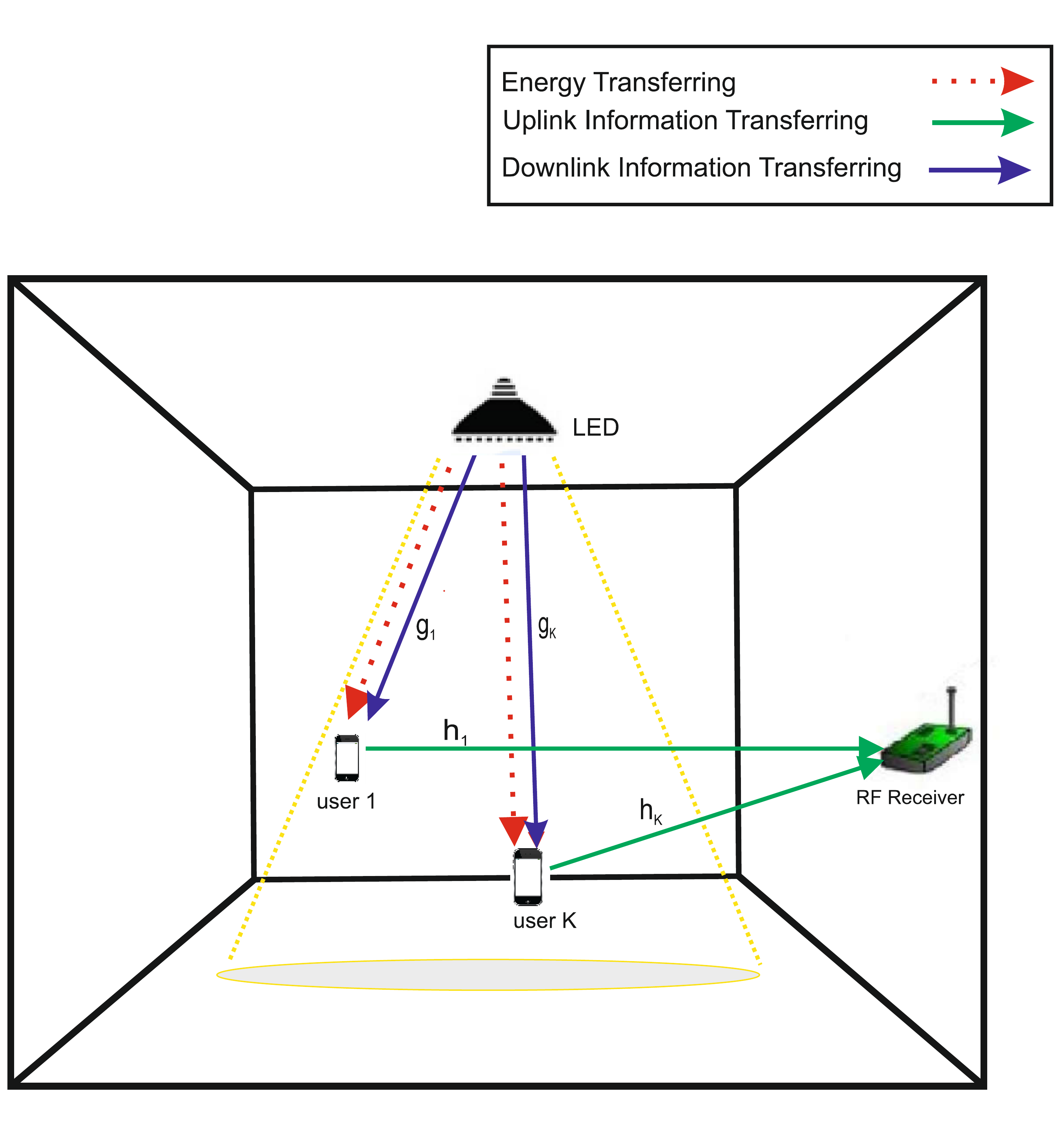}
		\caption{\footnotesize A SLIPT-based dual-hop VLC/RF system.}
	\end{figure}
	\section{system model}
	In this paper,~a VLC system is considered,~as shown in Fig.~1,~which consists of one LED as an access point (AP) that serves up to $K$ users from set $\mathcal{K}=\{1,~...,~K\}$.~The LED is placed at the center of the ceiling with a limited coverage area,~and all users are randomly located inside this area.~Moreover,~the LED transmitter supplies users with energy through time division multiple access (TDMA) scheme.~It is assumed that UL RF channels are quasi-static block fading,~where the channel coefficients vary from one to another but remain fixed during each block.~Also,~it is assumed that the RF receiver can obtain perfect channel state information (CSI){\cite{Zenaidi}}.~When user $k$ is served, the optical transmitted signal of the LED can be represented as  
	\begin{equation}\label{transmit signal}
	x_k=\sqrt{P_{\text{LED}}}s_k+a,\: k\in \mathcal{ K},
	\end{equation}    
	where $P_{\text{LED}}$ and $a$ are transmit power of the LED and DC-offset,~respectively.~Furthermore,~the peak amplitude of the $s_k$ is given by $-A \leq s_k \leq A$,~where $A$ can be obtained by modulation order.~In the VLC systems with IM/DD,~the transmit signal $x_k$, $\forall k$, should be non-negative,~so we have
	$P_{\text{LED}}\leq \frac{a^2}{A^2}.$
	Besides,~the average optical power of transmit signal $x_k$ can be written as
	\begin{equation}\label{transmit signal 2}
	\mathbb{E}(x_k)=\mathbb{E}(\sqrt{P_{\text{LED}}}s_k+a)=a.
	\end{equation}    
	Due to the radiant energy of the LED,~the value of DC-offset (average optical power) should be restricted,~i.e.,~$
	a\leq I_{\text{max}},$  
	where $I_{\text{max}}$ denotes the maximum average optical power.~At the $k$-th user,~the received optical power is converted into electrical current through direct detection at the PD.~The electrical current, $y_k$ at the output of the PD can be expressed as
	\begin{align}\label{received 2}
	&y_k= \xi g_k(\sqrt{P_{\text{LED}}}s_k+a)+z_k= \xi g_k\sqrt{P_{\text{LED}}}s_k+\xi g_ka+z_k\nonumber\\
	&= I_{{AC,k}}+ I_{{DC,k}}+z_k,\: k\in \mathcal{ K},
	\end{align}
	where $\xi$ is the optical to electrical conversion efficiency; $z_k$ is an additive white Gaussian noise (AWGN) coming from shot noise and thermal noise with total variance of $\delta^2=N_0 W$,~where $W$ is the modulation bandwidth,~and $N_0$ is the noise power spectral density.~More specifically,~the photocurrent consists of the DC and AC signals.~The AC component,~i.e.,~$P_{\text{LED}}$,~is utilized for decoding information,~while the DC component,~i.e.,~$a$ is used for harvesting energy.~In addition,~$g_k$ denotes the DL channel link between the LED and user $k$,~which is mainly provided by the LOS link.~Accordingly,~the DL channel link can be modeled as follows \cite{Komine}
	\begin{equation}\label{channel}
	{g_k} = \left\{ \begin{array}{l}\frac{{(m + 1)A_p{R_p}}}{{2\pi ({L^2} + {r_k^2})}}{\cos ^m}(\phi )T_s(\varphi)T_f(\varphi)\cos (\varphi ),{\rm{\quad  0}} \le \varphi  \le \psi ;{\rm{ }}\\0,{\rm{    \quad\quad\quad\quad\quad\quad\quad\quad\quad\quad\quad\quad\quad\quad\quad\quad\:\:    \text{otherwise},}}\end{array} \right.{\rm{ }}
	\end{equation}
	where $m=-1/{\log _2}(\cos ({\Phi _{\frac{1}{2}}}))$ is the Lambertian radiation.~$\phi$ and $\varphi$ represent the angles of irradiance and incidence between the LED and each user,~respectively.~$A_p$ and $R_p$ represent the detection area and responsivity of the PD,~respectively.~$\psi$ and $\Phi _{\frac{1}{2}}$ are the half of the solar panel’s field of view (FOV) and semi-angle of LED,~respectively.~The vertical and horizontal distance between the LED and each user are defined as $L$ and $r_k$,~respectively.~Moreover,~$T_s(\varphi)$ is the gain of the optical filter and $T_f(\varphi)$ is non-imaging concentrator of gain which is given by
	\begin{equation}
	{T_f}(\varphi )=
	\begin{cases}
	\frac{{{n^2}}}{{{{\sin }^2}\varphi }}, & 0, \le \varphi  \le \psi, \\
	0, & \text{otherwise,}\\
	\end{cases}       
	\end{equation}
	where $n$ is the refractive index.~It is notable that the UL channel link (i.e., the channel link between each user and the RF receiver) is denoted by $h_k$ throughout the paper.

	\section{Scenario A: WPT-based Dual-hop VLC/RF System}
	In this section,~we consider a simple scenario wherein all users transmit their own corresponding RF signals in the UL by utilizing the harvesting power in the DL,~as shown in Fig.~2.~In particular,~the LED sends a constant DC signal without any data,~and all users harvest energy from the signal transmitted by the LED.~The transmit signal of the LED is $x_k=a$, $ k \in \mathcal{K}$,~so the received signal at the $k$-th user has the form of $y_k= g_ka+z_k$.~By ignoring the received power of the noise,~the amount of received energy at each user can be expressed as
	$E_k={\eta Ta^2g_k^2}$,~where $\eta\in(0,1]$ is the conversion energy efficiency.~It should be noted that the time slot is normalized,~i.e.,~$T=1$,~which indicates that all users constantly harvest energy in the DL.~Then,~the transmit power of user $k$ in the UL can be stated as $p_{k}=\frac{\eta a^2g_k^2}{\tau_k^{ul}},$ where $\tau _k^{ul}$ is the portion of time given to user $k$ for transmitting its RF data in the UL.~The signal-to-noise ratio (SNR) at the RF receiver is given as $ \Gamma_k=\frac{{{p_{k}}{{\left| {{h_k}} \right|}^2}}}{{{\sigma ^2}}}$,~where $|h_k|^2$ is the channel power gain between each user and the RF receiver over the UL and it is assumed that $h_k$ follows a Rayleigh distribution with $\mathbb{E}[|h_k|^2]$ normalized to unity,~and $\sigma ^2$ denotes the received power noise at the RF receiver.~Consequently,~the UL sum-rate can be written as \cite{Lapidoth}
	\begin{equation}\label{sum1}
	R_{\text{sum}}^{ul}=\sum_{k\in\mathcal{K}} {\tau _k^{ul}{{\log }_2}(1 +\Gamma_k)}.
	\end{equation}
	\subsection{{Problem Formulation}}
	\begin{figure}
		\centering
		\includegraphics[width=0.4\textwidth]{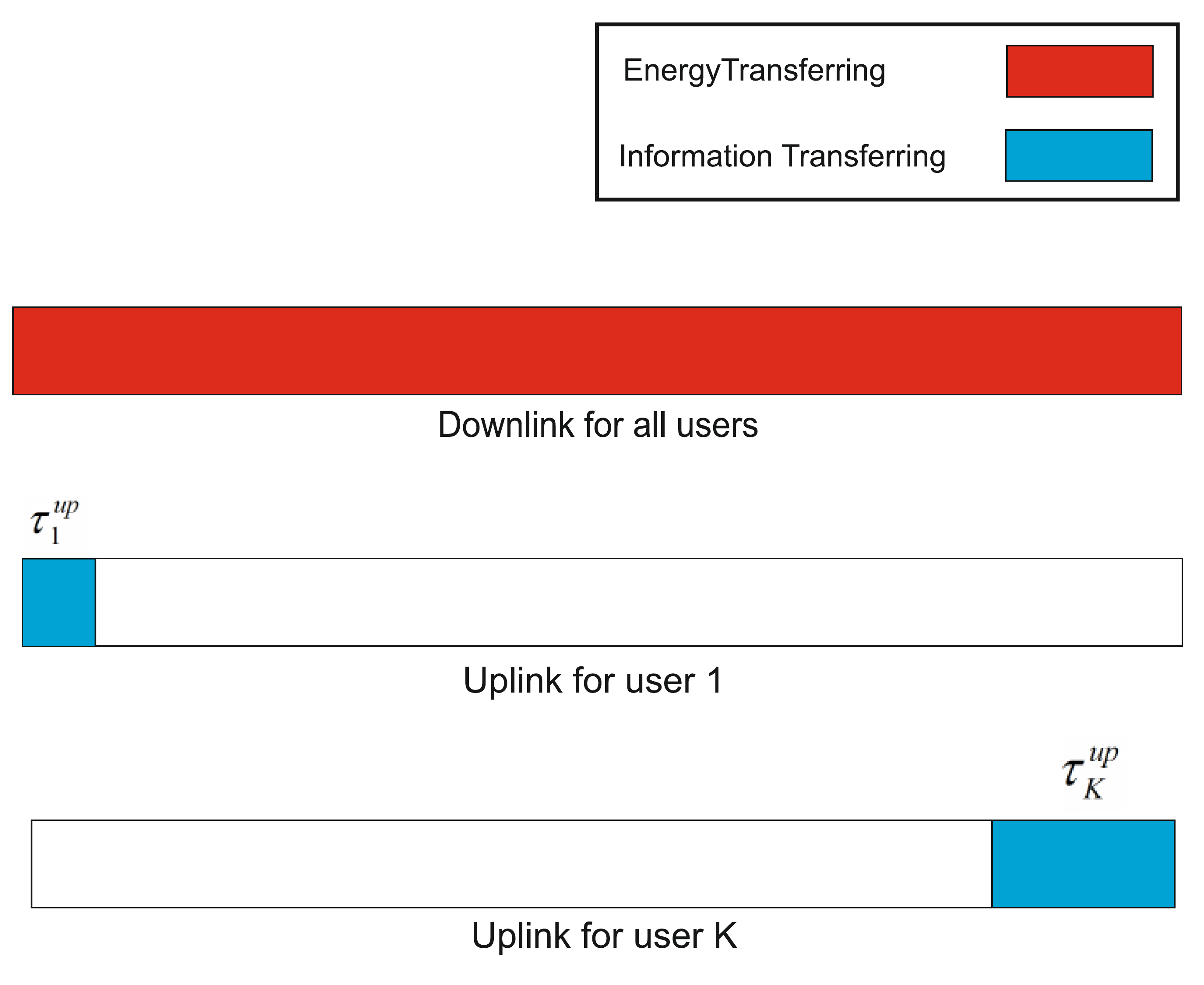}
		\caption{{\footnotesize Scenario A: Time frame of the WPT-based dual-hop VLC/RF system.}}
	\end{figure}
	In this subsection,~we aim at maximizing the sum-rate of the UL transmission subject to the total power constraint of the LED.~Thus,~the maximization problem can be mathematically formulated as
	\begin{subequations}
		\begin{align}
		& \text{P1}:  \underset{\tau _k^{ul}} {\text{max}} \: \sum_{k\in\mathcal{K}}  {\tau _k^{ul}{{\log }_2}(1 + \frac{{{\eta a^2g_k^2}{{\left| {{h_k}} \right|}^2}}}{{{\tau _k^{ul}\sigma ^2}}})}, \\
		&\text{s.t.} \quad \sum_{k\in\mathcal{K}}  {\tau _k^{ul}}\leq 1,\label{9b}\\
		&\quad\quad {a ^2}\leq p_{\text{max}},\label{9c}\\
		&\quad\quad \tau_k^{ul}\geq0,\:  k\in \mathcal{ K},\label{9e}
		\end{align} 
	\end{subequations}
	where (\ref{9b}) is the total time allocated to all $K$ users for the UL transmission.~Due to the luminous ability of LED device and eye safety,~the value of DC-offset $a$ should be limited which is expressed by (\ref{9c}),~$p_{\text{max}}$ denotes the maximum transmit power of the LED,~and (\ref{9e}) denote the UL time allocation.
	\begin{theorem}
		At the optimal solution of problem (P1),~It can be shown that constraints (\ref{9b}) and (\ref{9c}) hold with equality,~i.e.,~$\sum_{k\in\mathcal{K}}  {\tau _k^{ul}}=1$ and ${a^2}= p_{\text{max}}$.
	\end{theorem}
	
	\textit{Proof:}
	Assume that $\big\{(\tau _k^{ul})^*,a^*\big\}$ are the optimal solutions of problem (P1),~and the corresponding optimal value of  $R_{\text{sum}}^{ul}$ is $(R_{\text{sum}}^{ul})^*$.~This solution also satisfies $\sum_{k\in\mathcal{K}}  ({\tau _k^{ul}})^*<1$.~Then,~we create another new solution as $\{\bar{\tau} _k^{ul},\bar{a}\}$,~where $\bar{\tau} _k^{ul}=\alpha({\tau _k^{ul}})^*$ and $\bar{a}=a^*$.~Also,~we have $\alpha=
	1/({\sum_{k\in\mathcal{K}}  {({\tau _k^{ul}})^*}})>1$ such that $\sum_{k\in\mathcal{K}}  \bar{\tau} _k^{ul}=1$,~then the corresponding optimal sum-rate is $\bar{R}_{\text{sum}}^{ul}$.~In particular,~the solutions set $\{\bar{\tau} _k^{ul},\bar{a}\}$ satisfies all the constraints of problem (P1).~By substituting this solutions set into the main problem,~we obtain higher sum-rate since the objective function is an increasing function with respect to ${\tau _k^{ul}}$.~Hence,~it can be proved that ${\sum_{k\in\mathcal{K}} \tau _k^{ul}}=1$ by contradiction.~Similarly,~it can be shown that ${a^2}= p_{\text{max}}$ satisfies by contradiction.
	\subsection{{Optimal Solution}}
	Base on Theorem 1,~the equivalent form of problem (P1) can be given by
	\begin{subequations}
		\begin{align}
		& \text{P2}:  \underset{\tau _k^{ul}} {\text{max}} \: \sum_{k\in\mathcal{K}}  {\tau _k^{ul}{{\log }_2}(1 + \frac{{{\eta g_k^2}{{\left| {{h_k}} \right|}^2 p_{max}}}}{{{\tau _k^{ul}\sigma ^2}}})}, \\
		&\text{s.t.} \quad \sum_{k\in\mathcal{K}}  {\tau _k^{ul}}=1,\label{10b}\\
		&\quad\quad \tau_k^{ul}\geq0,\:  k\in \mathcal{ K}.
		\end{align} 
	\end{subequations}
	According to \cite{Boyd},~the perspective of function $f(x)$ is 
	$g(x,y)$ defined as $g(x,y)= yf(\frac{x}{y})$,~$g =\{(x,~y)\:|\:\frac{x}{y}\in f,~y> 0 \}$.~If $f(x)$ is a concave function,~then so is its perspective function $g(x,y)$.~Thus,~the transformed problem (P2) is a convex problem and satisfies the Slater’s condition.~Then,~the optimal solution can be achieved by applying the Lagrange dual method.~The Lagrangian function can be written as
	\begin{align}\label{Lagrangian 1}
	&\mathcal{L}= \sum_{k\in\mathcal{K}}  {\tau _k^{ul}{{\log }_2}(1 + \frac{{{\eta p_\text{max}g_k^2}{{\left| {{h_k}} \right|}^2}}}{{{\tau_k^{ul}\sigma ^2}}})}+ \lambda(1-\sum_{k\in\mathcal{K}}  {\tau _k^{ul}}),
	\end{align}
	where $\lambda$ is the dual Lagrange multiplier
	associated with constraint (\ref{10b}).~Therefore,~the dual problem can be expressed as:
	\begin{equation}\label{derivative 2}
	\mathop {\min }\limits_{\lambda} {\rm{ }}\mathop {\max }\limits_{\tau_k^{ul}}\:\mathcal{L}(\tau_k^{ul},\lambda).
	\end{equation}
	The optimal value of $(\tau_k^{ul})^*$ can be obtained by solving the following equation
	\begin{equation}\label{derivative12}
	\frac{{\partial\mathcal{L}}}{{\partial \tau _k^{ul}}} = {\log _2}(1 + \frac{{y_k}}{\tau_k^{ul}}) - \frac{{{y_k}}}{{(\tau_k^{ul} + {y_k})\ln2}} - \lambda=0,\:  k\in \mathcal{ K},
	\end{equation}
	where $y_k= \frac{{{\eta g_k^2}{{\left| {{h_k}} \right|}^2p_{max}}}}{{{\sigma ^2}}}$.~By indicating the inverse function of $f(x) = \log (x) + \frac{1}{x\ln2}-1$ as $f^{-1}(x)$,~equation (\ref{derivative12}) can be written as:
	\begin{equation}\label{tau}
	\tau_k^{ul}=\left.\frac{y_k}{f^{-1}(\lambda)-1}\right|_0^{{1}},\:  k\in \mathcal{ K},
	\end{equation}
	where $x|^b_a = \text{min}\big\{\text{max}\{x,~a\},~b\big\}$.~Since the dual function is differentiable,~we adopt the gradient-based method \cite{Bertsekas} to find the dual variables as follows
	\begin{equation}\label{variable1}
	\lambda(t+1)=\big[\lambda(t)-\alpha(t)(1-\sum_{k\in\mathcal{K}}  {\tau _k^{ul}})\big]^+,
	\end{equation}
	where $\alpha(t)\geq0$ is a positive step size and $[x]^+=\text{max}\{x,0\}$.~Finally,~the Lagrange multiplier in
	(\ref{variable1}) and the primal variable $\tau_k^{ul}$ can be optimized in an iterative manner to solve the main problem.
	\section{Scenario B: SLIPT-based dual-hop VLC/RF system}
	In this scenario,~both UL and DL schemes are considered as shown in Fig.~3.~Let $T$ denotes the frame transmission time which is divided into $K$ orthogonal time slots given by $\tau_k^{\text{dl}}$ and $\tau_k^{\text{ul}}$,~$ k \in \mathcal{ K}$,~respectively.~{For simplicity},~we consider that $T=1$ second.~For the DL transmission,~each user receives data from the AC component in the portion of $\tau_k^{\text{dl}}$ and harvests energy from the DC component in the portion of $1-\tau_k^{\text{dl}}$.~Then,~each user utilizes harvested energy to \begin{figure}
		\centering
		\includegraphics[width=0.4\textwidth]{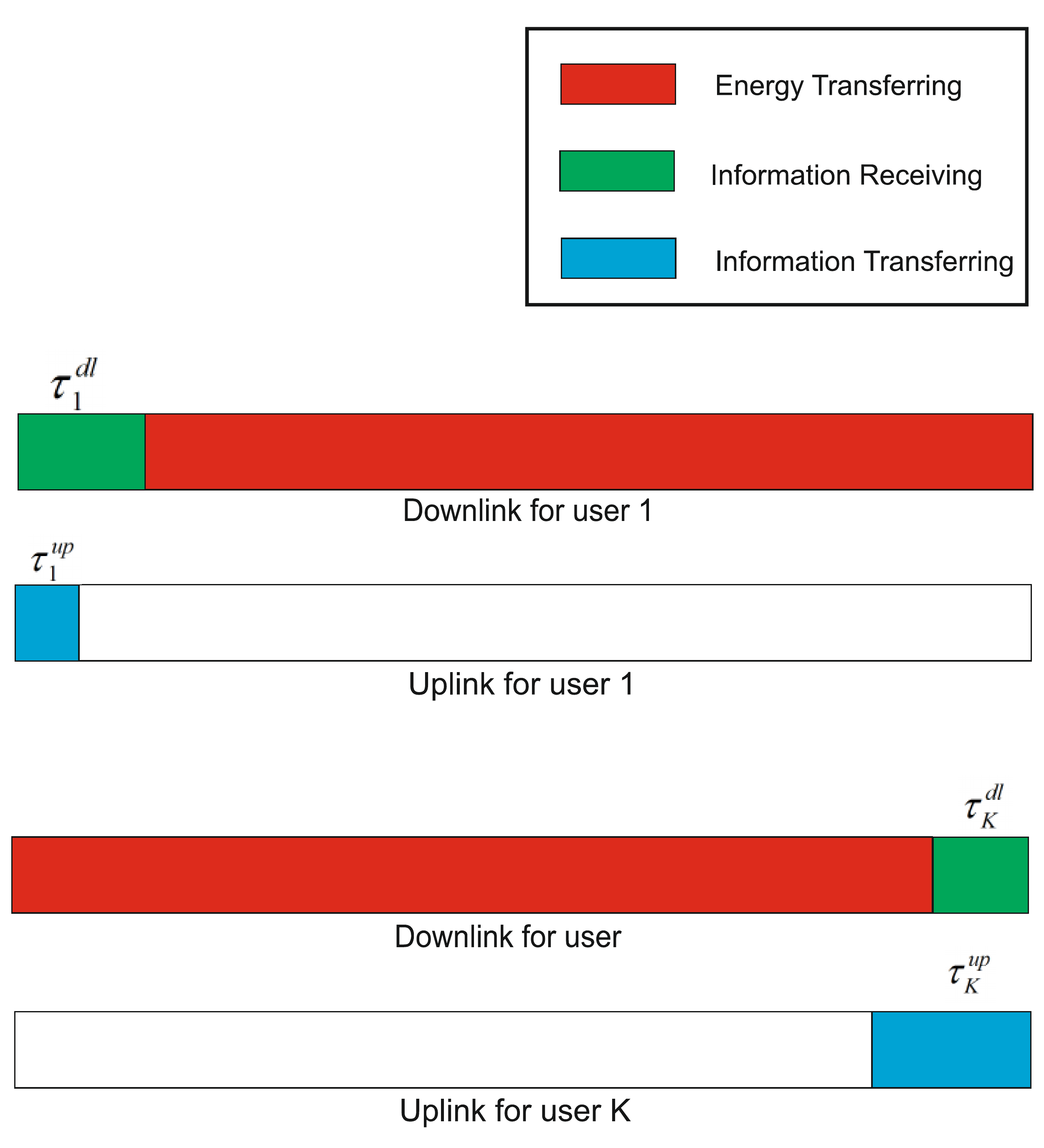}
		\caption{{\footnotesize Scenario B: time frame of the SLIPT-based dual-hop VLC/RF system.}}
	\end{figure} send its RF signal in the $\tau^{ul}_k$ second.~It is notable that the frequency division duplex (FDD) scheme is adopted for the UL and DL.~Consequently,~the achievable lower bound of DL data rate for the user $k$ and sum-rate are given by
	\begin{subequations}\begin{align}
		&R_k^{dl}=\tau _k^{dl}{{\log }_2}(1 + \frac{e}{2\pi}\frac{{{P_{k,\text{LED}}}{g_k^2}}}{{{\delta^2}}}),\\
		&R_{\text{sum}}^{\text{dl}}=\sum_{k\in\mathcal{K}} \tau _k^{dl}{{\log }_2}(1 + \frac{e}{2\pi}\frac{{{P_{k,\text{LED}}}}{g_k^2}}{{{\delta^2}}}),
		\end{align}
	\end{subequations}
	respectively.~Accordingly,~the amount of harvested energy in the DL and the transmit power of each user in the UL can be written as
	\begin{align}\label{energy dl 2}
	&E_k=\eta{a^2}g_k^2(1-\tau_k^{dl}),
	&{p}_{k}=\frac{\eta{a^2}g_k^2(1-\tau_k^{dl})}{\tau_k^{ul}},\: k\in \mathcal{ K},
	\end{align}
	respectively.~Subsequently,~the UL sum-rate can be stated as
	\begin{equation}\label{SNR up 2}
	R_{\text{sum}}^{\text{ul}}=\sum_{k\in\mathcal{K}} \tau_k^{ul}\log_2(1+\frac{ { p}_{k}|h_k|^2}{\sigma^2}),
	\end{equation}
	where $|h_k|^2$ denotes the channel power gain between each user and the RF receiver over the UL,~and it is assumed that $h_k$ follows a Rayleigh distribution,~and $\sigma ^2$ describes the received power noise at the RF receiver.	
	\subsection{Problem Formulation}
	{In dual-hop VLC/RF systems,~UL/DL sum data rates maximization problems are both desirable system design objectives.~Hence,~we first introduce two single-objective optimization problems concerning these desirable objective functions.~Then,~we combine them into a multi-objective optimization problem (MOOP).~Under the framework of MOOP,~a performance trade-off can be established.~The first considered problem is designed to maximize the total DL sum-rate which is given by}

	\textit{\textbf{Problem 1: DL sum-rate Maximization:}}
	\begin{subequations}
		\begin{align}
		& \text{P1}:  \underset{\tau _k^{ul},\tau_k^{dl},P_{k,\text{LED}}} {\text{maximize}} \: \sum_{k\in\mathcal{K}} \tau _k^{dl}{{\log }_2}(1 + \frac{e}{2\pi}\frac{{{ P_{k,\text{LED}}}}{g_k^2}}{{{\delta^2}}}),\\
		&\text{s.t.} \quad \sum_{k\in\mathcal{K}}  {\tau _k^{ul}}\leq 1,\quad\quad \sum_{k\in\mathcal{K}}  {\tau _k^{dl}}\leq 1,\label{P1-1}\\
		&\quad\quad \sum_{k\in\mathcal{K}} ({a^2 +P_{k,\text{LED}}}) \tau_k^{dl}\leq p_{\text{max}},\label{P1-2}\\
		&\quad\quad  \sum_{ i \ne k,~k\in \mathcal{K}} \eta{a^2 }g^2_k\tau_i^{dl}\geq e_{k,\text{min}},~ k\in \mathcal{ K},\label{P11-3} \\
		&\quad\quad P_{k,\text{LED}}\leq \frac{a^2}{A^2},\:\: 0\leq a\leq I_{\text{max}},\:  k\in \mathcal{ K},\label{P1-3}\\
		&\quad\quad \tau_k^{ul}\geq0,\: \tau_k^{dl}\geq0,\:  k\in \mathcal{ K}.\label{P1-4}
		\end{align} 
	\end{subequations}
	In (P1),~the total DL sum-rate is maximized by jointly optimizing DL time allocation $\tau _k^{dl}$,~$ k\in \mathcal{ K}$,~UL time allocation $\tau _k^{up}$,~$ k\in \mathcal{ K}$,~and LED transmit power $P_{k,\text{LED}}$,~$ k\in \mathcal{ K}$.~(\ref{P1-1}) enforces the total time allocation constraint for both UL and DL transmissions.~(\ref{P1-2}) is the total transmit power restriction of the LED,~where $p_{\text{max}}$ indicates the maximum transmit power.~(\ref{P11-3}) is the total harvested energy by user $k$ and $e_{k,\text{min}}$ denotes the minimum harvested energy.~(\ref{P1-3}) and (\ref{P1-4}) denote the non-negativity of transmit power as well as UL and DL time allocations.
	
	Similarly,~for the UL sum-rate maximization system design,~we have the same constraints as for problem (P1).~Therefore~the problem formulation for UL is given by
	
	\textit{\textbf{Problem 2: UL sum-rate Maximization:}}
	\begin{subequations}
		\begin{align}
		& \text{P2}:  \underset{\tau _k^{ul},\tau_k^{dl},P_{k,\text{LED}}} {\text{maximize}}\: \sum_{k\in\mathcal{K}} \tau_k^{ul}\log_2(1+\frac{p_k|h_k|^2}{\sigma^2}),\\
		&\text{s.t.} \quad \text{(\ref{P1-1})-(\ref{P1-4})},
		\end{align} 
	\end{subequations}
	where $p_{k},$ $\forall k\in \mathcal{ K}$ is based on (\ref{energy dl 2}).
	The relation between the objectives of (P1) and (P2) is prominent in the dual-hop VLC/RF system.~A nontrivial trade-off between these two objective functions typically occurs in the dual-hop VLC/RF system.~Hence,~we require to design a resource allocation algorithm by taking into account these two system design objectives.~For this purpose,~we adopt MOOP to address this issue.~In general,~MOOP is a mathematical framework that investigates the trade-off between these two conflicting objective functions,~which yields Pareto optimal set\cite{Multi-Objective}.
	
	In the following,~we introduce a MOOP in order to strike a balance between the DL and UL sum-rates by jointly optimizing the power allocation,~DL time allocation,~and UL time allocation.~Thus,~the MOOP formulation can be expressed as follows.
	
	\textit{\textbf{Problem 3: MOOP:}}
	\begin{subequations}
		\begin{align}
		& \text{P3}:  \underset{\tau _k^{ul},\tau _k^{dl},P_{k,\text{LED}}} {\text{minimize}} \: -\sum_{k\in\mathcal{K}} \tau _k^{dl}{{\log }_2}(1 +\frac{e}{2\pi} \frac{{{P_{k,\text{LED}}}}{g_k^2}}{{{\delta^2}}}),\\
		&\quad\quad \underset{\tau _k^{ul},\tau _k^{dl},P_{k,\text{LED}}} {\text{minimize}}  -\sum_{k\in\mathcal{K}} \tau_k^{ul}\log_2(1+\frac{p_k|h_k|^2}{\sigma^2}),\\
		&\text{s.t.} \quad \text{(\ref{P1-1})-(\ref{P1-4})}.
		\end{align} 
	\end{subequations}
	
	The weighted Tchebycheff approach can be adopted in order to address the MOOP by employing a set of predefined weights \cite{Multi-Objective}-\cite{Multi-Objective3}.~Specifically,~this approach is able to provide a complete Pareto optimal set even though the MOOP is not-convex.~For the sake of simplification,~we express the objective functions of (P1) and (P2) as $G_1(\mathcal{F})$ and $G_2(\mathcal{F})$,~respectively,~where $\{\tau _k^{up},\tau _k^{dl},P_{k,\text{LED}}\}\in \mathcal{F}$ denotes the set of optimization variables.~Thus,~the optimization problem based on the weighted Tchebycheff method can be written as
	\begin{subequations}
		\begin{align}
		& \text{P4}:  \underset{\mathcal{F}} {\text{min}} \:\: \underset{i=\{1,2\}} {\text{max}} \: \left\{\lambda_i \left(G_i(\mathcal{F})-G^*_i(\mathcal{F})\right)\right\}, \\
		&\text{s.t.} \quad \text{(\ref{P1-1})-(\ref{P1-4})},
		\end{align} 
	\end{subequations}
	where $G^*_i$ is the optimal value with respect to problem $i\in\{1,2\}$,~i.e.,~$G_1(\mathcal{F})=-\sum_{k\in\mathcal{K}} \tau _k^{dl}{{\log }_2}(1 +\frac{e}{2\pi} \frac{{{ P_{k,\text{LED}}}}{g_k^2}}{{{\delta^2}}})$ and $G_2(\mathcal{F})=-\sum_{k\in\mathcal{K}} \tau_k^{ul}\log_2(1+\frac{p_k|h_k|^2}{\sigma^2})$.~Also,~$\lambda_i$ is a constant predefined parameter enforced on the $i$-th objective function such that $0\leq\lambda_i\leq1$ and $\sum_i\lambda_i=1$.~In particular,~various predefined parameters can be adopted by solving problem (P4) to obtain resource allocation policies.
	\subsection{Optimal Solutions of Problem 1 	and 2}
	According to Theorem 1,~constraints (\ref{P1-1}) and (\ref{P1-2}) are held with equality which can be similarly proved by contradiction,~i.e.,~we have 
	\begin{align}
	&\sum_{k\in\mathcal{K}}  {\tau _k^{dl}}=1,\quad\quad\sum_{k\in\mathcal{K}}  {\tau _k^{ul}}=1,\label{key}\\
	&\sum_{k\in\mathcal{K}} ({a^2 +P_{k,\text{LED}}}) \tau_k^{dl}= p_{\text{max}}.\label{key1}
	\end{align}
	\textit{\textbf{Optimal Solution of Problem 1}}:
	Problem (P1) can be rewritten as
	\begin{subequations}
		\begin{align}
		& \text{P1-1}:  \underset{\mathcal{F}} {\text{max}} \: \sum_{k\in\mathcal{K}} \tau _k^{dl}{{\log }_2}(1 + \frac{e}{2\pi}\frac{{{ P_{k,\text{LED}}}}{g_k^2}}{{{\delta^2}}}),\\
		&\text{s.t.} \quad \text{(\ref{key}),~(\ref{key1}),~(\ref{P11-3})-(\ref{P1-4})},
		\end{align} 
	\end{subequations}
	which is a non-convex optimization problem due to the existence of coupling between variables.~To handle it,~we introduce a set of new auxiliary variables as $b_k=a^2\tau_k^{dl}$ and $\tilde{P}_{k,\text{LED}}=P_{k,\text{LED}}\tau_k^{dl}$,~$ k\in \mathcal{ K}$.~Then,~problem (P1-1) can be recast as
	\begin{subequations}
		\begin{align}
		& \text{P1-2}:  \underset{\tau _k^{ul},\tau_k^{dl},\tilde{P}_{k,\text{LED}},b_k} {\text{maximize}} \: \sum_{k\in\mathcal{K}} \tau _k^{dl}{{\log }_2}(1 + \frac{{{\tilde{P}_{k,\text{LED}}}}}{{{\tau_k^{dl}}}}\gamma_k),\\
		&\text{s.t.} \quad \text{(\ref{key}),~(\ref{P1-4})},\label{27d}\\
		&\quad\quad \sum_{k\in\mathcal{K}} {\tilde{P}_{k,\text{LED}}+b_k}= p_{\text{max}},\label{P4-2}\\
		&\quad\quad  \sum_{ i \ne k,~k\in \mathcal{K}}\eta b_kg^2_k\geq e_{k,\text{min}},~ k\in \mathcal{K},\label{enrgy}\\
		&\quad\quad {\tilde{P}_{k,\text{LED}}}\leq \frac{b_k}{A^2},\:\: b_k\leq I^2_{\text{max}}\tau_k^{dl},\:  k\in \mathcal{K},\label{P4-3}
		\end{align} 
	\end{subequations}
	where $\gamma_k=\frac{e g_k^2}{2\pi\delta^2}$.~It is observed that problem (P1-2) is convex and the optimal solutions can be obtained by the standard interior-point method \cite{Boyd}.~However,~the complexity of the typical interior-point method is high.~Thus,~we propose a low-complexity iterative approach with two steps to obtain optimal solutions.~To provide more insights into the solution design,~we apply the Karush-Kuhn-Tucker (KKT) conditions \cite{Boyd}.~Accordingly,~the Lagrangian function of problem (P1-2) can be written as
	\begin{align}\label{Lagrangian 2}
	&\mathcal{L}= \sum_{k\in\mathcal{K}} \tau _k^{dl}{{\log }_2}(1 + \frac{{{\tilde{P}_{k,\text{LED}}}}}{{{\tau_k^{dl}}}}\gamma_k)+ X,
	\end{align}
	where 
	\begin{align}\label{29}
	&X=\lambda(1-\sum_{k\in\mathcal{K}}  {\tau _k^{ul}})+\zeta(1-\sum_{k\in\mathcal{K}}  {\tau _k^{dl}})+\sum_{k\in \mathcal{K}}\omega(I^2_{\text{max}}\tau_k^{dl}-b_k)\nonumber\\&+\psi(\sum_{k\in\mathcal{K}} {\tilde{P}_{k,\text{LED}}+b_k}- p_{\text{max}})+\sum_{k\in\mathcal{K}}\nu_k(\frac{b_k}{A^2}-\tilde{P}_{k,\text{LED}})\nonumber\\&+\sum_{k\in \mathcal{K}}\mu_k\big(\sum_{ i \ne k,~k\in \mathcal{K}}\eta b_kg^2_k-e_{k,\text{min}}\big),
	\end{align}
	where $\lambda$,~$\zeta$,~$\psi$,~$\nu_k$,~$\omega$ and $\mu_k$ are the non-negative Lagrange multipliers associated with the constraints of problem (P1-2).~By taking the partial derivative of $\mathcal{L}$ with respect to $\tau _k^{up}$,~$\tau_k^{dl}$,~$b_k$ and $\tilde{P}_{k,\text{LED}}$,~respectively,~we have
	\begin{align}
	&\frac{\partial\mathcal{L}}{\partial\tau _k^{up}}=-\lambda,\label{28}\\
	&\frac{\partial\mathcal{L}}{\partial\tau _k^{dl}}={{\log }_2}(1 + \frac{{{\tilde{P}_{k,\text{LED}}}}}{{{\tau_k^{dl}}}}\gamma_k)-\frac{{{\tilde{P}_{k,\text{LED}}}}{\gamma_k}}{\tau_k^{dl}+{{\tilde{P}_{k,\text{LED}}}}{\gamma_k}}-\zeta+w_kI^2_{\text{max}},\label{30}\\
	&\frac{\partial\mathcal{L}}{\partial{b_k}}=\psi+\frac{\nu_k}{A^2}-w_k+\sum_{ i \ne k,~k\in \mathcal{K}}\eta\mu_i g^2_k,\\
	&\frac{\partial\mathcal{L}}{\partial{\tilde{P}_{k,\text{LED}}}}=\psi-\nu_k-\frac{{\tau^{dl}_k\gamma_k}}{\tau_k^{dl}+{{\tilde{P}_{k,\text{LED}}}}{\gamma_k}}\label{32}.
	\end{align}
	To guarantee that $\frac{\partial\mathcal{L}}{\partial\tau _k^{up}}$ is bounded above,~we should have $\frac{\partial\mathcal{L}}{\partial\tau _k^{up}}\leq 0$.~On the other hand,~$\frac{\partial\mathcal{L}}{\partial\tau _k^{up}}< 0$ leads to $\tau _k^{up}=0$,~while $\frac{\partial\mathcal{L}}{\partial\tau _k^{up}}=0$ results in $\tau _k^{up}\geq 0$.~Consequently,~we obtain
	\begin{align}
	\tau _k^{up}\left\{ \begin{array}{l}\in (0,1],\quad\:\:\:{\rm{   if \: \:  \frac{\partial\mathcal{L}}{\partial\tau _k^{up}}= 0,}}\\= 0,\quad\quad\quad\:{\rm{    if \:\:   \frac{\partial\mathcal{L}}{\partial\tau _k^{up}}< 0.~}}\end{array} \right.
	\end{align}
	However,~it is hard to obtain a
	closed-form expressions of $(\tau _k^{dl},\tilde{P}_{k,\text{LED}},b_k)$.~Hence,~we solve problem (P1-2) for fixed $\tilde{P}_{k,\text{LED}}$ to obtain $(\tau _k^{dl},b_k)$ and then we obtain $\tilde{P}_{k,\text{LED}}$ with given $(\tau _k^{dl},b_k)$ iteratively.~It is notable that since optimizing $(\tau _k^{dl},b_k)$ and updating $\tilde{P}_{k,\text{LED}}$ both  aim to increase the objective function of (P1-2),~the overall algorithm is ensured to converge.~Next,~equation (\ref{30}) can be stated as
	\begin{align}
	\tau _k^{dl}=\frac{\tilde{P}_{k,\text{LED}}\gamma_k}{f^{-1}(\zeta-w_kI^2_{\text{max}})-1}\bigg|^1_0,~ k\in \mathcal{K},
	\end{align}
	where the inverse function $f^{-1}(x)$ was defined in the first scenario.~Nevertheless,~the optimal value of $b_k$ cannot be obtained instantly since the Lagrange function is linear with respect to $b_k$.~To handle it,~we rewrite the objective function of (P1-2) based on the penalty function as follows
	\begin{subequations}
		\begin{align}
		& \text{P1-3}: \mathop {\text{max} }\limits_{\scriptstyle \tau _k^{up},\tau_k^{dl}\atop\scriptstyle \tilde{P}_{k,\text{LED}},b_k} \: \sum_{k\in\mathcal{K}} \tau _k^{dl}{{\log }_2}(1 +\frac{{{\gamma_k\tilde{P}_{k,\text{LED}}}}}{{{\tau_k^{dl}}}})+\alpha\log(1+b_k),\\
		&\text{s.t.} \quad \text{(\ref{27d})-(\ref{P4-3})}.
		\end{align} 
	\end{subequations}
	Problem (P1-3) is a convex problem,~and the optimal solutions almost equal to the main problem (P1-2) by choosing a small value of $\alpha$.~The optimal value of $b_k$ can be obtained by applying the KKT conditions as follows
	\begin{equation}
	b_k=\bigg[\frac{\alpha}{w_k-\psi-\frac{\nu_k}{A^2}-\sum\limits_{ k\in\mathcal{K},~i \ne k}\eta\mu_i g^2_k}-1\bigg]\bigg|^{p_{max}}_0,~ k\in \mathcal{K}.
	\end{equation}
	Then,~the dual variables are updated based on the gradient-based approach with given optimization variables \cite{Bertsekas}.~The updated dual variables are given by
	\begin{align}
	&\lambda(t+1)=\bigg[\lambda(t)-\beta(t)\big(1-\sum_{k\in\mathcal{K}}  {\tau _k^{ul}}\big)\bigg]^+,\label{38}\\
	&\zeta(t+1)=\bigg[\zeta(t)-\beta(t)\big(1-\sum_{k\in\mathcal{K}}  {\tau _k^{dl}}\big)\bigg]^+,\\
	&\psi(t+1)=\bigg[\psi(t)-\beta(t)\big(\sum_{k\in\mathcal{K}} {\tilde{P}_{k,\text{LED}}+b_k}- p_{\text{max}}\big)\bigg]^+,\label{c}\\
	&\nu_k(t+1)=\bigg[\nu_k(t)-\beta(t)\big(\frac{b_k}{A^2}-\tilde{P}_{k,\text{LED}}\big)\bigg]^+,\label{nu}\\
	&\omega(t+1)=\bigg[\omega(t)-\beta(t)\big(I^2_{\text{max}}\tau_k^{dl}-b_k\big)\bigg]^+,\\
	&\mu_k(t+1)=\bigg[\mu_k(t)-\beta(t)\big(\sum_{\scriptstyle k \in\mathcal{ K}\atop\scriptstyle i \ne k} \eta b_kg^2_k-e_{k,\text{min}}\big)\bigg]^+,~\label{43}
	\end{align}where $\beta(t)\geq0$ is a positive step size.~The dual variables $(\lambda,\zeta,\psi,\nu_k,\omega,\mu_k)$ and the primal variables $(b_k,\tau_k^{dl},\tilde{P}_{k,\text{LED}})$ can be optimized iteratively.~In the second step,~the optimal value of $\tilde{P}_{k,\text{LED}}$ can be obtained with fixed $(\tau _k^{dl},\tau_k^{ul},b_k)$.~From (\ref{32}),~we have
	\begin{align}\label{42}
	\tilde{P}_{k,\text{LED}}=\tau^{dl}_k\left[\frac{1}{\gamma_k}-\frac{1}{\psi-\nu_k }\right]^{p_{max}}_0,~ k\in \mathcal{K}.
	\end{align}
	As stated in (\ref{42}),~dual variables $(\psi,\nu_k)$ are required to obtain $\tilde{P}_{k,\text{LED}}$.~After calculating  $\tilde{P}_{k,\text{LED}}$ from (\ref{42}),~the gradient based method is exploited to update dual variables $(\psi,\nu_k)$ from (\ref{c}) and (\ref{nu}).~Since problem (P1-2) is convex with given $(\tau _k^{ul},\tau_k^{dl},b_k)$,~optimizing $\tilde{P}_{k,\text{LED}}$ and dual variables $(\psi,\nu_k)$ yields the optimal $\tilde{P}_{k,\text{LED}}$ iteratively.
	
	\textit{\textbf{Optimal Solution of Problem 2}}:
	Similarly,~problem (P2) can be restated as follows
	\begin{subequations}
		\begin{align}
		& \text{P2-1}: \underset{\tau _k^{ul},\tau_k^{dl},b_k,\tilde{P}_{k,\text{LED}}} {\text{maximization}}\: \sum_{k\in\mathcal{K}} \tau_k^{ul}\log_2(1+\frac{\bar{\gamma}_ka^2}{\tau_k^{ul}}-\frac{\bar{\gamma}_kb_k}{\tau_k^{ul}}),\\
		&\text{s.t.} \quad \text{(\ref{27d})-(\ref{P4-3}),}
		\end{align} 
	\end{subequations}
	where $\bar{\gamma}_k=\frac{\eta|h_k|^2g_k^2}{\sigma^2}$,~$\forall k$.~By exploiting the KKT conditions we obtain
	\begin{align}\label{Lagrangian 3}
	&\mathcal{L}= \sum_{k \in \mathcal{K}}\tau_k^{ul}\log_2(1+\frac{\bar{\gamma}_ka^2}{\tau_k^{ul}}-\frac{\bar{\gamma}_kb_k}{\tau_k^{ul}})+ X.
	\end{align}
	
	\begin {table}
	\caption{Default Simulation Parameters}
	\centering
	\begin {tabular}{|l|l|}
	\hline
	Parameter,~Notation                             & Value                         \\ \hline
	LED bandwidth,~$B$                        & $30$ MHz                         \\ \hline
	The physical area of a PD,~$A_p$                   & $1\text{cm}^2$                          \\ \hline
	Semi-angle of LED,~$\Phi_{1/2}$          & $60^o$                              \\ \hline
	The angle of irradiance,~$\phi$                  & $60^o$                     \\ \hline
	The angle of incidence,~$\varphi$                     & $45^o$                      \\ \hline
	Optical to electrical conversion efficiency & $0.53$ A/W                      \\ \hline
	Gain of optical filter,~$T_s$                               & $1 $                            \\ \hline
	Refractive index of PD lens                     & $1.5$                           \\ \hline
	Room size,~$(\text{L}\times\text{W}\times \text{H})$                               & $(5\text{m}\times5\text{m}\times3\text{m})$                     \\ \hline
	LED coordinate $(x_\text{LED},y_\text{LED},z_\text{LED})$                                 & $(2.5\text{m},2.5\text{m},3\text{m})$                \\ \hline
	RF receiver coordinate $(x_\text{RF},y_\text{RF},z_\text{RF}) $                               & $(4\text{m},1\text{m},1\text{m})$                    \\ \hline
	User height,~$h_\text{user}$                                     & $1$m                            \\ \hline
	Conversion energy efficiency,~$\eta$                              & $0.2$                          \\ \hline
	Optical to electrical conversion efficiency,~$\xi$                              & $1$                          \\ \hline
	\end {tabular}
	\end {table}
	In the following,~we take partial derivative of $\mathcal{L}$ with respect to $\tau_k^{dl}$,~$\tau _k^{up}$,~$b_k$ and $\tilde{P}_{k,\text{LED}}$,~respectively,~as follows
	\begin{align}
	&\frac{\partial\mathcal{L}}{\partial{\tau} _k^{dl}}=\omega I^2_{\text{max}}-\zeta,\\
	&\frac{\partial\mathcal{L}}{\partial\tau _k^{up}}=\log_2(1+\frac{\bar{\gamma}_k(a^2-b_k)}{\tau_k^{ul}})-\frac{\bar{\gamma}_k(a^2_k-b_k)}{\tau^{up}_k+\bar{\gamma}_k(a^2_k-b_k)}-\lambda,\label{44}\\
	&\frac{\partial\mathcal{L}}{\partial{b_k}}=-\frac{\tau^{up}_k\bar{\gamma}_k}{\tau^{up}_k+\bar{\gamma}_k(a^2_k-b_k)}+\psi+\frac{\nu_k}{A^2}-\omega+              \sum_{ i \ne k,~k\in \mathcal{K}}\eta\mu_i g^2_k,\label{49}\\
	&\frac{\partial\mathcal{L}}{\partial{\tilde{P}_{k,\text{LED}}}}=\psi-\nu_k.
	\end{align}	
	An iterative algorithm can be employed to solve problem (P2-1) by first fixing $b_k$ to obtain $\tau _k^{up}$ and vice versa.~To make sure that $\frac{\partial\mathcal{L}}{\partial{{\tau} _k^{dl}}}$ is bounded above,~we obtain
	\begin{align}
	{{\tau} _k^{dl}}\left\{ \begin{array}{l}\in (0,1],\quad\quad\quad{\rm{   if \:  \frac{\partial\mathcal{L}}{\partial{{\tau} _k^{dl}}}= 0,}}\\= 0,\quad\quad\quad\quad\:\:{\rm{    if \:  \frac{\partial\mathcal{L}}{\partial{{\tau} _k^{dl}}}< 0.}}\end{array} \right.
	\end{align}
	Similarly,~by setting $\frac{\partial\mathcal{L}}{\partial \tilde{P}_{k,\text{LED}}}=0$,~we have 
	\begin{align}
	{\tilde{P}_{k,\text{LED}}}\left\{ \begin{array}{l}\in (0,\frac{b_k}{A^2}],\quad\quad{\rm{   if \:  \frac{\partial\mathcal{L}}{\partial \tilde{P}_{k,\text{LED}}}= 0,}}\\= 0,\quad\quad\quad\quad\:{\rm{    if \:  \frac{\partial\mathcal{L}}{\partial \tilde{P}_{k,\text{LED}}}< 0.~}}\end{array} \right.
	\end{align}
	A close-form expression for ${\tau}_k^{dl}$ can be obtained as 
	\begin{align}
	\tau _k^{up}=\frac{(a^2_k-b^2_k)\gamma}{f^{-1}(\lambda)-1}\bigg|^1_0,~ k\in \mathcal{K},
	\end{align}
	where the inverse function $f^{-1}(x)$ was defined in the first scenario.~The dual optimization parameters can be updated from (\ref{38})-(\ref{43}).~Now,~for fixed ${\tau}_k^{up}$,~the optimal value of $b_k$ can be obtained from (\ref{49}) as follows
	\begin{align}
	b_k=\tau^{up}_k\bigg[\frac{1}{\bar{\gamma}_k}+\frac{1}{\omega-\psi-\frac{\nu_k}{A^2}-\sum\limits_{i \ne k,~k\in \mathcal{K}}\eta\mu_i g^2_k}\bigg]+a^2\bigg|^1_0,~ k\in \mathcal{ K}.
	\end{align}
	Similar to (P1-3),~problem (P2-1) can be solved iteratively by guaranteeing the convergence.
	
	\textit{\textbf{Optimal Solution of Problem 3}}: Problem (P3) can be transformed into its equivalent form as follows
	\begin{subequations}
		\begin{align}
		& \text{P3-1}:  \underset{\tau _k^{ul},\tau _k^{dl},P_{k,\text{LED}},t} {\text{minimize}} \:\:t \\
		&\quad\quad \lambda_i \big(G_i(\tau _k^{up},\tau _k^{dl},P_{k,\text{LED}})-G^*_i\big)\leq t,~\forall i\in\{1,2\},\\
		&\text{s.t.} \quad \text{(\ref{27d})-(\ref{P4-3}),}
		\end{align} 
	\end{subequations}
	where $t$ is an auxiliary optimization variable.~We note that (P3-1) is a convex optimization problem that can be solved efficiently by using CVX \cite{cvx}.
	\begin{figure}
		\centering
		\includegraphics[width=0.5\textwidth]{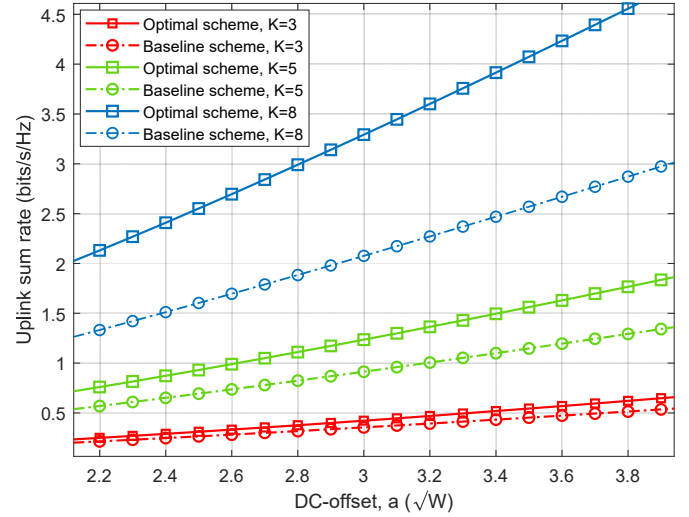}
		\caption{\small Uplink sum-rate versus DC-offset with different number of users and resource allocation (scenario A).}
	\end{figure}
	
	\begin{figure}
		\centering
		\includegraphics[width=0.495\textwidth]{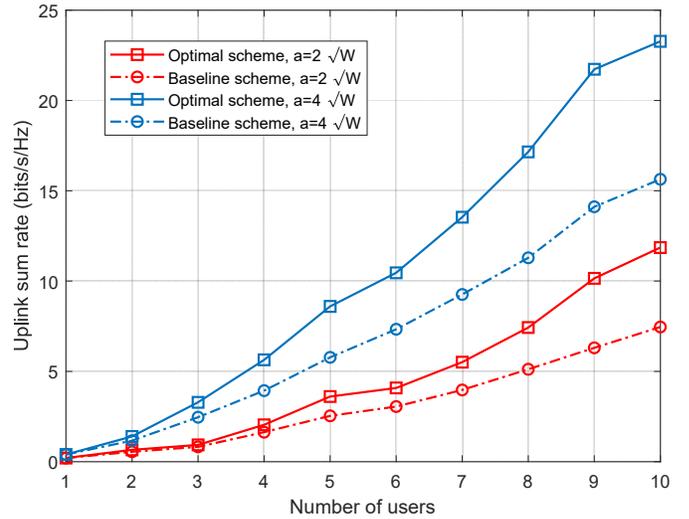}
		\caption{\small Uplink sum-rate versus number of users with different DC-offsets (scenario A).}
	\end{figure}
	\section{Numerical Results}
	In this section,~we present numerical results and simulation to evaluate and compare the proposed scenarios with the baseline schemes.~In each scenario,~we compare the proposed optimal time allocation with the baseline schemes.~In the baseline scheme,~each user is assigned with equivalent and fixed time duration.~Simulation parameters and their corresponding values are given in Table I \cite{Rakia1,Rakia2}. Users' locations are generated uniformly 500 times in simulations.~We adopt the path loss model for RF as $d^{-\alpha}$,~where $\alpha=1.8$ is the path loss exponent (typically,~$\alpha$ takes values 1.6--1.8 for indoor situations \cite{Rappaport}).~Moreover,~the UL channels (RF channel links) are assumed to follow the Rayleigh fading model.

	\begin{figure}
		\centering
		\includegraphics[width=0.5\textwidth]{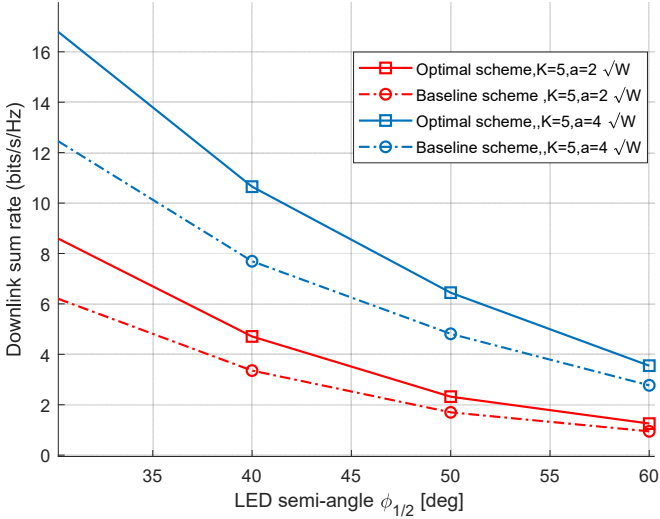}
		\caption{\small Uplink sum-rate versus LED semi-angle with different number of users,~DC-offset (scenario A).}
	\end{figure}
	\begin{figure}
		\centering
		\includegraphics[width=0.5\textwidth]{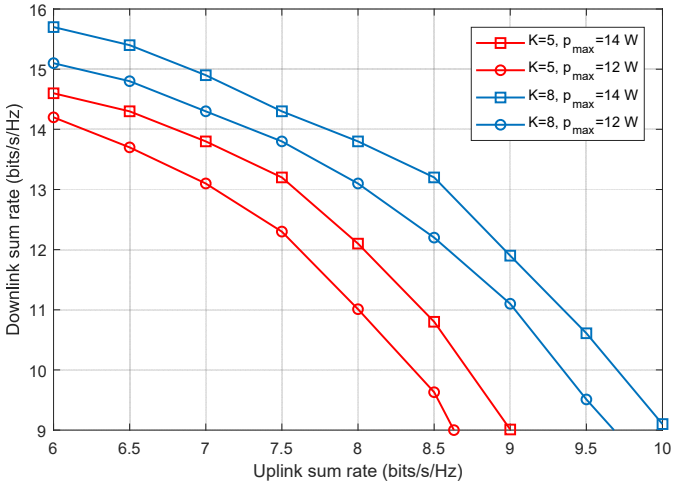}
		\caption{\small System performance trade-off region between the downlink sum-rate and the uplink sum-rate for $a=2\sqrt{{W}}$ and $e_{k,\text{min}}=-20$ dBm with different number of users and transmit power schemes (scenario B).}
	\end{figure}
	\subsection{Scenario A: WPT-based dual-hop VLC/RF system}
	
	Fig.~4 displays the total UL sum-rate versus different values of DC-offset for a different number of users.~Moreover,~the performance of both optimal resource allocation and baseline schemes are compared.~For both schemes,~the UL sum-rate increases by increasing the DC-offset value since each user harvest more energy from the DC component of the transmit signal,~which leads to a higher transmission power in the UL.~It can be found that the proposed optimal resource allocation always outperforms the baseline schemes for a different number of users,~especially when the DC-offset increases.

	Figs.~5 shows the UL sum-rate versus the number of users.~From this figure,~we can observe that the UL sum-rate monotonically increases by increasing the number of users.~As DC-offset, $a$ grows the UL sum-rate increases since more energy can be harvested in the DL.~Also,~the optimal scheme outperforms the baseline schemes,~which arise from the fact that each transmission time allocated to a user with a better channel condition.

	Fig.~6 shows the DL sum-rate versus the LED semi-angle for a different number of users and DC-offsets.~It is observed that increasing the LED semi-angle results in a decreasing the DL sum-rate.~Increasing the LED semi-angle reduces the harvested energy since channel gain decreases and therefore,~the benefit of receiving the beam of the LED is eliminated.
	
	\subsection{Scenario B: SLIPT-based dual-hop VLC/RF system}

	Fig.~7 exhibits the trade-off region for the DL and UL sum-rates with a different number of users and power allocation schemes when the minimum required harvested energy is $-20$ dBm.~The trade-off region in this figure is obtained by solving problem (P3-1) through changing the values of $0\leq\lambda_i\leq1$, $\forall \lambda_i \in \{1, 2\},$ uniformly.~It can be easily observed that the DL sum-rate is a monotonically decreasing function versus the UL sum-rate.~This means that the DL and UL sum-rate maximization problems are conflicting system design objectives.~In other terms,~a resource allocation algorithm maximizing the DL sum-rate is unable to maximize the UL sum-rate simultaneously in the considered dual-hop VLC/RF system.~As expected,~the trade-off region expands as the number of users increases,~i.e.,~it achieves a bigger trade-off region since more users participate in the ID and EH process.~Besides,~with a fixed number of users,~as the transmit power budget of LED increases each user can decode more information and harvest more energy,~which would be used for UL transmission.~This process accordingly leads to the improvement of the trade-off region.~As a matter of fact,~for DL transmission,~each user can decode information and harvest energy for the transmission time of other users,~which depends on the radiated power.~Hence,~more transmit power promotes decoding information and harvesting energy.
	

	%
	
	\section{Conclusion}

	In this paper,~we proposed two scenarios in a dual-hop VLC/RF system where both scenarios include UL and DL transmissions.~For the first scenario,~all users harvest energy from the received lightwave (DC component) continuously in all time slots over DL and then use the harvested energy to transmit information signals in the UL over the RF channels based on TDMA scheme.~The proposed optimization framework has concentrated on the proper choice of the DC-offset.~For the second scenario,~the LED transmits information (AC component) and power (DC component) simultaneously in the DL based on the TS-based SLIPT in contrast to the first scenario.~In each UE,~the harvested energy is exploited to transmit information signals in the UL over the RF channels.~A MOOP framework was proposed to strike a trade-off between EH and information decoding over DL and UL transmission with a focus on the proper selection of the transmit power of the LED,~UL and DL transmission times.~The presented numerical results have confirmed that the proposed scenarios considerably improve the sum data rate as compared to the fixed policies.

\end{document}